\PassOptionsToPackage{unicode}{hyperref}
\PassOptionsToPackage{dvipsnames,svgnames*,x11names*}{xcolor}

\documentclass[]{article}

\usepackage{arxiv}

\usepackage[utf8]{inputenc} 
\usepackage[T1]{fontenc}    
\usepackage{lmodern}        
\usepackage{xcolor}
\usepackage{hyperref}       
\usepackage{url}            
\usepackage{booktabs}       
\usepackage{amsfonts}       
\usepackage{nicefrac}       
\usepackage{microtype}      
\usepackage{graphicx}
\usepackage{setspace}
\usepackage{orcidlink}
\usepackage{fontspec}

\hypersetup{
  pdftitle={Quantifying Inconsistent Mediation: The Mediated Magnitude Share and the Degree of Suppression},
  pdfkeywords={mediation analysis, inconsistent mediation, proportion mediated, mediated magnitude share, directional consistency index},
  breaklinks=true,
  bookmarks=true,
  colorlinks=true,
  linkcolor={magenta},
  filecolor={Maroon},
  citecolor={blue},
  urlcolor={Blue},
  pdfcreator={LaTeX via pandoc}}

\title{Quantifying Inconsistent Mediation: The Mediated Magnitude Share and the
Degree of Suppression}

\author{
    Md. Niamul Islam Sium%
  \hspace{0.12em}\orcidlink{0009-0005-1822-8294}%
  \\
  University of Texas at El Paso \\
  Texas, USA \\
  \texttt{\href{mailto:msium@miners.utep.edu}{\nolinkurl{msium@miners.utep.edu}}} \And
  Shafayet Khan Shafee%
  \hspace{0.12em}\orcidlink{0009-0002-8021-3788}%
  \thanks{corresponding author}%
  \\
  Independent Researcher \\
  Dhaka, Bangladesh \\
  \texttt{\href{mailto:sshafee@isrt.ac.bd}{\nolinkurl{sshafee@isrt.ac.bd}}} \And
  Mohammad Hridoy Patwary%
  \hspace{0.12em}\orcidlink{0009-0008-4920-7017}%
  \\
  International Centre for Diarrhoeal \\ Disease Research, Bangladesh (icddr,b) \\
  Dhaka, Bangladesh \\
  \texttt{\href{mailto:mpatwary@isrt.ac.bd}{\nolinkurl{mpatwary@isrt.ac.bd}}} \And
  Bishal Sarker%
  \hspace{0.12em}\orcidlink{0009-0001-0812-4142}%
  \\
  University of Texas at El Paso \\
  Texas, USA \\
  \texttt{\href{mailto:bsarker3@miners.utep.edu}{\nolinkurl{bsarker3@miners.utep.edu}}}}

\usepackage[round,sort]{natbib}

\usepackage{float}
\usepackage{caption}
\DeclareCaptionFont{tenpt}{\fontsize{10}{12}\selectfont}
\usepackage{mathtools}
\usepackage{booktabs}
\usepackage{array}
\usepackage{doi}
\usepackage{url}
\usepackage{multirow}
\usepackage{threeparttable}
\begin{document}
\maketitle

\begin{abstract}
The proportion mediated (PM), the indirect effect divided by the total effect,
is intuitive when direct and indirect effects act in the same direction. Under
inconsistent mediation, however, opposing pathways partially cancel, so PM can
fall outside the unit interval and becomes unstable as the total effect
approaches zero. We propose two complementary effect-scale measures: the
mediated magnitude share (MMS), which quantifies the absolute magnitude of an
indirect pathway relative to the combined absolute magnitudes of the direct and
indirect pathways, and the degree of suppression, which quantifies the fraction
of this combined magnitude cancelled by opposing pathways. Both measures are
bounded between 0 and 1, and MMS reduces to PM under consistent mediation. We
extend both measures to multiple indirect pathways and, through a logarithmic
transformation, to binary outcomes on ratio scales. In simulations with
continuous and binary outcomes, PM estimates were highly dispersed when the
total effect was small relative to the pathway magnitudes, whereas estimates of
MMS and the degree of suppression remained bounded. We illustrate the measures
using previously published estimates from the Adolescents Training and Learning
to Avoid Steroids (ATLAS) program, where PM was \(-0.30\), MMS was \(0.19\), and the
degree of suppression was \(0.38\), showing how the two measures describe the
relative pathway magnitudes and their cancellation under inconsistent mediation.
\end{abstract}


\setstretch{1.12}

\section{Introduction}\label{sec-01}

Mediation analysis decomposes the total effect of a treatment on an outcome into
two components: the indirect effect (IE), which operates through the mediator,
and the direct effect (DE), which operates independently of the mediator.
Together, these effects aggregate to the total effect (TE), such that
\(\mathrm{TE} = \mathrm{DE} + \mathrm{IE}\). When the DE and IE have the same sign,
the proportion mediated (PM), \(\mathrm{PM} = \mathrm{IE} / \mathrm{TE}\), provides
a convenient summary of the mediator's contribution to the total effect. Under
this form of consistent mediation, PM lies between 0 and 1 and can be interpreted
as the fraction of the TE attributable to the IE \citep{alwin1975, mackinnon2008}.

However, this interpretation becomes flawed when the DE and IE have opposite
signs, a situation commonly referred to as inconsistent mediation or
suppression \citep{mackinnon1995, mackinnon2000}. In this setting, the DE and IE
partially cancel each other, so that the magnitude of the TE is smaller than
the sum of their magnitudes, that is,
\(\lvert\mathrm{TE}\rvert=\lvert\mathrm{DE}+\mathrm{IE}\rvert<\lvert\mathrm{DE}\rvert+\lvert\mathrm{IE}\rvert\).
Consequently, since PM uses the TE as its denominator, it can take values outside
the \([0,1]\) range and become arbitrarily large in magnitude as the TE approaches
zero \citep{vanderweele2015, mackinnon1995, preacher2011, cheung2009}. In the case of
perfect suppression, \(\mathrm{DE} = -\mathrm{IE}\), the TE is zero and PM is
undefined, even though both the DE and IE can be substantively large.

\begin{figure}[!ht]
\centering
\includegraphics[width=0.55\textwidth]{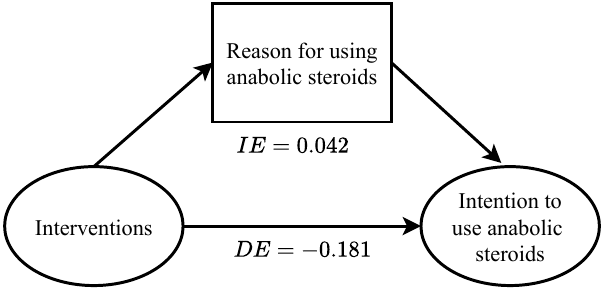}
\caption{Path diagram illustrating the inconsistent mediational pathway in the
case of Adolescents Training and Learning to Avoid Steroids (ATLAS) program. The
intervention affects the outcome (intentions to use anabolic steroids) both
directly and through the mediator (reasons for using anabolic steroids). The
estimated indirect and direct effects are $\mathrm{IE} = 0.042$ and 
$\mathrm{DE} = -0.181$, respectively \citep{mackinnon2000}.}
\label{fig:atlas-dag}
\end{figure}

A real example of inconsistent mediation comes from the Adolescents Training and
Learning to Avoid Steroids (ATLAS) program, a randomized trial evaluating a
multidimensional intervention to reduce anabolic steroid use among high school
football players \citep{goldberg1996}. This study was used to illustrate inconsistent
mediation, focusing on the mediational pathway through reasons for using
anabolic steroids \citep{mackinnon2000}, as shown in Figure \ref{fig:atlas-dag}.
The intervention was effective overall, reducing participants' intentions to use
steroids, with an estimated TE of \(-0.139\). However, the intervention also increased
the number of reasons for using anabolic steroids, which in turn was associated
with increased intentions to use steroids. Thus, this indirect pathway operated
against the overall intervention effect, with an estimated IE of \(0.042\), whereas
the corresponding DE was \(-0.181\). The PM is therefore,
\(\mathrm{PM}=0.042/(-0.139)\approx-0.30\). Although this value is mathematically
well-defined, it does not admit a straightforward proportion interpretation: it
cannot be logically interpreted as the fraction of the TE attributable to the
indirect pathway when the indirect pathway operates in the opposite direction
from the TE.

Therefore, the issue is not merely that PM can produce values outside its nominal
range, but also that its denominator does not provide an appropriate reference under
suppression. A measure of mediated contribution should instead use a reference
denominator that remains well-defined under both consistent and inconsistent
mediation and ensures a value bounded within the unit interval, thereby retaining
a proportional interpretation. At the same time, the extent of suppression should
be quantified separately, since the contribution of a pathway and the extent to
which pathways counteract one another represent distinct features of a mediation
system.

Several existing approaches have sought to characterize mediation without relying
directly on the PM. For example, variance-accounted measures summarize mediation
through changes in explained variance \citep{fairchild2009, deHeus2012, lachowicz2018}.
These measures address mediation from a variance-explanation perspective rather
than by partitioning the TE into the DE and IE on the effect scale. Therefore,
they do not directly answer the question of what proportion of the TE is attributable
to the indirect pathway. Thus, there remains a need for an effect-scale measure
that retains the intuitive proportional interpretation of PM while remaining bounded
and well-defined under inconsistent mediation.

We address this gap by introducing two complementary measures: the Mediated
Magnitude Share (MMS) and the degree of suppression. MMS quantifies the relative
magnitude of the IE to the combined magnitudes of the DE and IE, providing a
bounded and stable measure of mediated contribution. Importantly, MMS reduces
exactly to PM under consistent mediation; it therefore generalizes rather than
replaces the conventional PM measure. The degree of suppression quantifies the
proportion of the combined magnitude of DE and IE that is cancelled when the DE
and IE have opposing signs. The two measures capture distinct features of the
same inconsistent mediation system: MMS describes how the combined magnitude is
distributed between the indirect and direct pathways, whereas the degree of
suppression quantifies how much of that magnitude is cancelled by their opposing
directions. Together, these two measures provide an interpretable characterization
of inconsistent mediation.

We first develop these measures for the two-pathway setting of one direct and one
indirect pathway (Section \ref{sec-02}). We then extend our discussion to the
multiple-pathway case, including multiple mediators (Section \ref{sec-03}), and
consider its application to binary outcomes on the ratio scale (Section \ref{sec-04}).
We subsequently examine the finite-sample behavior of the proposed measures through
simulation studies (Section \ref{sec-05}) and illustrate their application using
the ATLAS example (Section \ref{sec-06}). Finally, we conclude with a general
discussion (Section \ref{sec-07}).

\section{The two-pathway case}\label{sec-02}

Consider a mediation system with one direct and one indirect pathway; we define
\(d=\lvert\mathrm{DE}\rvert\), \(m=\lvert\mathrm{IE}\rvert\), and \(L=d+m\). We assume
\(L>0\), excluding the case in which both the DE and IE are zero. While
\(\lvert\mathrm{TE}\rvert\) corresponds to the magnitude of the sum of DE and IE,
the quantity \(L\) represents the sum of their magnitudes. By the triangle inequality,
\begin{equation}
\lvert\mathrm{TE}\rvert = \lvert\mathrm{DE} + \mathrm{IE}\rvert 
\le \lvert\mathrm{DE}\rvert+\lvert\mathrm{IE}\rvert = L \;.
\label{eq:treq}
\end{equation}
Equality holds when DE and IE have the same sign, corresponding to consistent
mediation, whereas the inequality is strict when they have opposite signs,
corresponding to inconsistent mediation.

\subsection{Mediated Magnitude Share}\label{mediated-magnitude-share}

\begin{figure}[!htbp]
  \centering
  \includegraphics[width=0.9\textwidth]{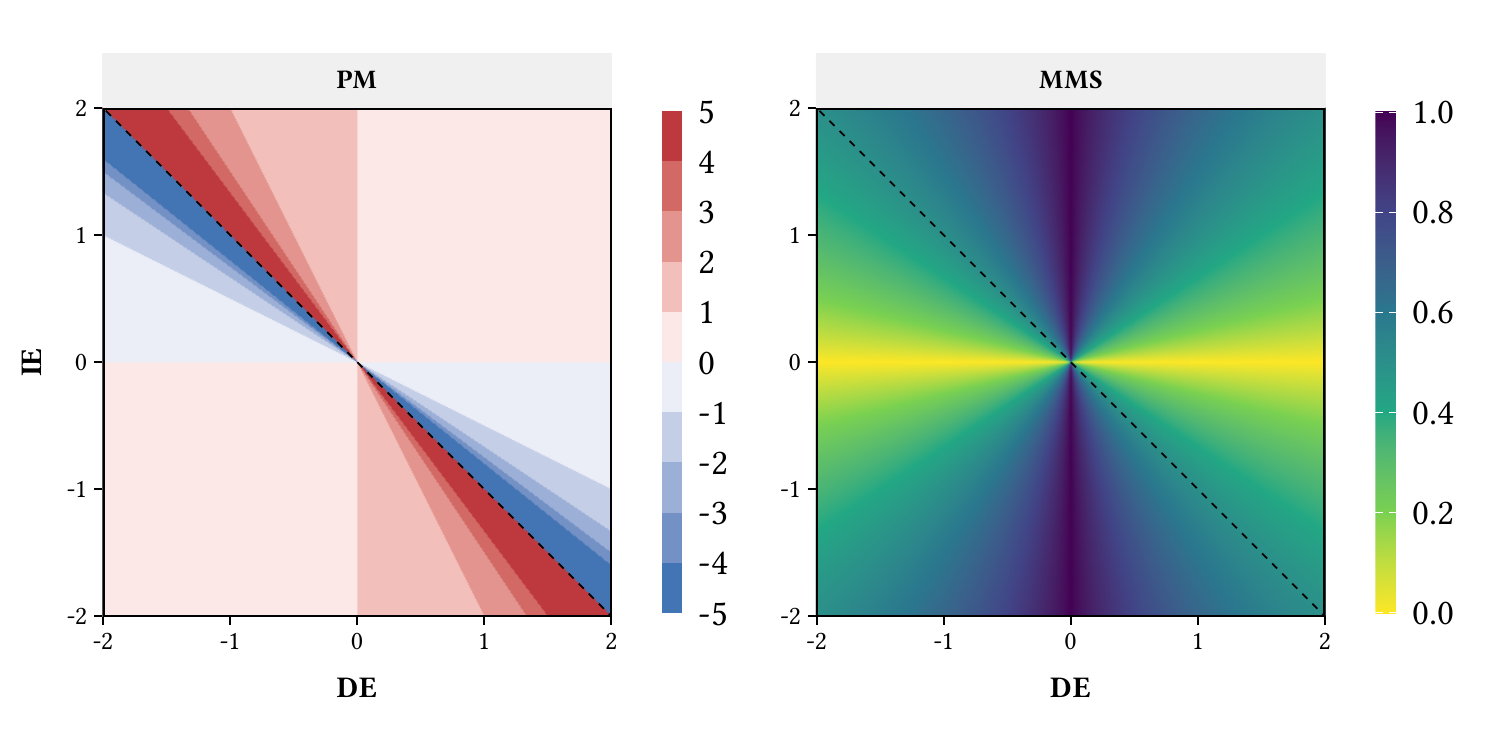}
  \caption{Behavior of the proportion mediated (PM, left) and the mediated 
    magnitude share (MMS, right) over the plane of direct effects (DE, 
    horizontal axis) and indirect effects (IE, vertical axis), each ranging 
    from $-2$ to $2$. The total effect is $\mathrm{TE}=\mathrm{DE}+\mathrm{IE}$. 
    Because PM is unbounded and diverges to $\pm\infty$ as $\mathrm{TE}\rightarrow0$, 
    PM values were truncated to the range $[-5, 5]$ for illustration purposes 
    only; MMS is shown without truncation, as it is bounded to the unit interval 
    by construction. The dashed diagonal denotes $\mathrm{TE}=0$, along which PM 
    is undefined; MMS remains well-defined there (equal to $0.5$) except at the 
    origin, where both DE and IE are zero.}
  \label{fig:heatmap}
\end{figure}

\begin{figure}[!htbp]
  \centering
  \includegraphics[width=0.95\textwidth]{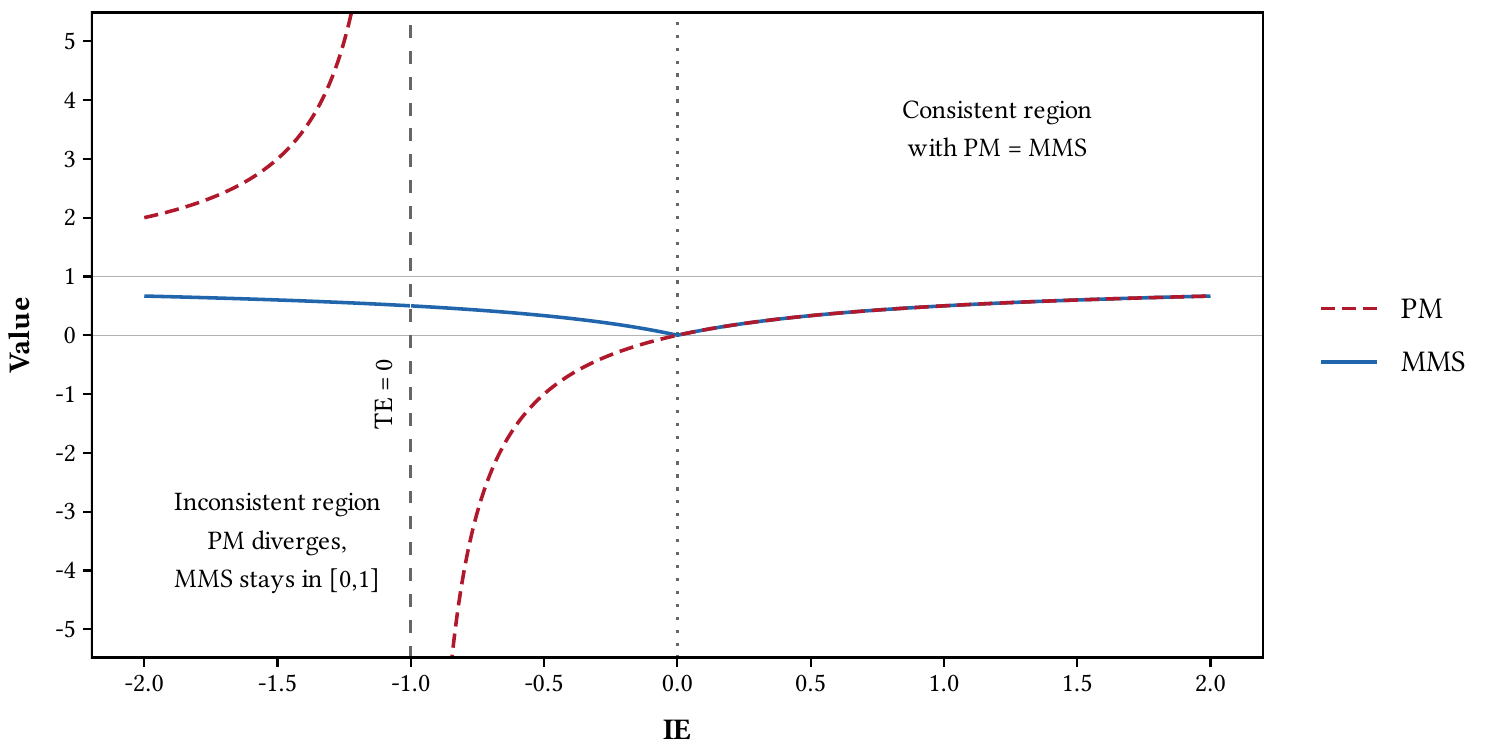}
  \caption{Proportion mediated (PM, red dashed) and mediated magnitude share 
    (MMS, blue solid) as functions of the indirect effect (IE), with the direct 
    effect fixed at $\mathrm{DE}=1$. Both measures are unchanged when DE and IE 
    are multiplied by a common positive constant or both reversed in sign, so 
    the choice $\mathrm{DE}=1$ entails no loss of generality. Because 
    $\mathrm{DE}>0$, the sign of IE determines the mediation regime: 
    $\mathrm{IE}>0$ corresponds to consistent mediation, where 
    $\mathrm{PM}=\mathrm{MMS}$, and $\mathrm{IE}<0$ to inconsistent mediation. 
    The vertical dotted line at $\mathrm{IE}=0$ marks the boundary between the 
    two regimes. The vertical dashed line at $\mathrm{IE}=-\mathrm{DE}=-1$ marks 
    perfect suppression, where the total effect is zero and PM is undefined; PM 
    is therefore drawn as two separate branches. As IE approaches $-1$, PM 
    diverges to $-\infty$ from the right and to $+\infty$ from the left, whereas 
    MMS remains within $[0,1]$ throughout and equals $0.5$ at the singularity. 
    Horizontal gray lines mark $0$ and $1$, the bounds of MMS. The vertical axis 
    is truncated to $[-5,5]$ for legibility.}
  \label{fig:trajectory}
\end{figure}

MMS measures the share of the combined magnitude of DE and IE that is
attributable to the magnitude of IE. Formally, we define
\begin{equation}
\mathrm{MMS}
= \frac{\lvert\mathrm{IE}\rvert}{\lvert\mathrm{DE}\rvert + \lvert\mathrm{IE}\rvert}
= \frac{m}{L} \;.
\label{eq:mms}
\end{equation}
Since \(0\leq m\leq L\), it follows that \(0 \le \mathrm{MMS} \le 1\), with
\(\mathrm{MMS}=0\) if and only if \(\mathrm{IE}=0\), and \(\mathrm{MMS}=1\) if and only
if \(\mathrm{DE}=0\). Under consistent mediation, TE has the same sign as DE and IE,
and therefore, from Equation \ref{eq:treq} and \ref{eq:mms}, it follows that
\begin{equation*}
\mathrm{MMS} 
= \frac{\lvert \mathrm{IE} \rvert}{\lvert \mathrm{TE} \rvert} 
= \frac{\mathrm{IE}}{\mathrm{TE}} = \mathrm{PM} \;.
\end{equation*}
Figure \ref{fig:heatmap} illustrates the distinction between MMS and PM over the
joint space of DE and IE. In the same-sign regions, the two measures coincide;
when the pathways oppose one another, PM becomes increasingly unstable as
\(\mathrm{TE} \rightarrow 0\), whereas MMS remains within \([0,1]\). Figure
\ref{fig:trajectory} shows the same transition along a one-dimensional path from
consistent to inconsistent mediation.

\subsection{Degree of Suppression}\label{degree-of-suppression}

The degree of suppression quantifies the extent to which the direct and indirect
pathways cancel one another. It is defined as the proportion of the combined
magnitude of DE and IE that is cancelled by their opposing directions:
\begin{equation}
\sigma
= 1-\frac{\lvert\mathrm{TE}\rvert}{\lvert\mathrm{DE}\rvert+\lvert\mathrm{IE}\rvert}
= 1-\frac{\lvert\mathrm{TE}\rvert}{L} \;.
\label{eq:sigma}
\end{equation}
Since \(L > 0\), Equation \ref{eq:treq} implies that \(\sigma \in [0,1]\). Under
consistent mediation, DE and IE have the same sign, so \(\lvert\mathrm{TE}\rvert=L\)
and \(\sigma=0\). Under inconsistent mediation, their opposing signs produce
suppression, with \(0<\sigma<1\) for partial suppression and \(\sigma=1\) under
complete suppression when \(\mathrm{TE}=0\). Thus, \(\sigma\) provides a scale-free
measure of the proportion of the combined magnitude of DE and IE that is lost
through their opposing directions.

\section{Multiple pathways}\label{sec-03}

Consider a mediation system with one direct pathway and \(k>1\) indirect
pathways operating through parallel mediators, that is, mediators that are each
affected by the treatment but not by one another. The TE is then decomposed as
\begin{equation*}
\mathrm{TE}=\mathrm{DE}+\sum_{i=1}^{k}\mathrm{IE}_i \;,
\end{equation*}
where \(\mathrm{IE}_i\) \((i = 1,\ldots,k)\) denotes the IE through the \(i\)th
mediator. The two-pathway case considered in Section \ref{sec-02} corresponds to
\(k=1\). With multiple pathways, suppression can occur among any subset of the
direct and indirect pathways. In particular, two or more indirect pathways may
operate in opposite directions and partially or completely offset one another.
Consequently, the magnitude of the TE may substantially understate the magnitudes
of the individual pathways, just as in the two-pathway case, but the sources of
suppression are more varied.

\subsection{Magnitude shares and the degree of suppression}\label{magnitude-shares-and-the-degree-of-suppression}

Define the combined magnitude of the direct and indirect pathways as
\(L = \lvert\mathrm{DE}\rvert + \sum_{i=1}^{k}\lvert\mathrm{IE}_i\rvert \;\).
Assuming \(L>0\), the MMS of indirect pathway \(i\) is
\begin{equation*}
\mathrm{MMS}_i = 
\frac{\lvert\mathrm{IE}_i\rvert}{L},
\quad i=1,\ldots,k \;.
\end{equation*}
Thus, \(\mathrm{MMS}_i\in[0,1]\) for \(i=1,\ldots,k\). The MMS of an indirect
pathway quantifies its share of the combined magnitude of DE and all IE,
irrespective of their signs. When all nonzero effects have the same sign,
there is no suppression and \(\lvert\mathrm{TE}\rvert=L\). In this case,
\begin{equation*}
\mathrm{MMS}_i
= \frac{\lvert\mathrm{IE}_i\rvert}{\lvert\mathrm{TE}\rvert}
= \frac{\mathrm{IE}_i}{\mathrm{TE}}\;,
\end{equation*}
so each MMS reduces to the corresponding pathway-specific proportion of the TE.
However, when pathways have opposing signs, \(L>\lvert\mathrm{TE}\rvert\), and the
MMS remains bounded, whereas the corresponding signed proportion
\(\mathrm{IE}_i/\mathrm{TE}\) does not. The definition and interpretation of the
degree of suppression remain unchanged in the multiple-pathway setting: as in
the two-pathway case (Equation \ref{eq:sigma}), \(\sigma\) quantifies the
proportion of the combined magnitude that is suppressed by opposing directions:
\begin{equation*}
\sigma
= 1-\frac{\lvert\mathrm{TE}\rvert}{L} \;.
\end{equation*}

\subsection{Indirect-versus-indirect suppression}\label{indirect-versus-indirect-suppression}

With multiple indirect pathways, suppression need not arise from opposition
between the DE and IE. Two or more indirect pathways may themselves operate in
opposite directions and substantially offset one another. This
``indirect-versus-indirect'' suppression is not possible in the single-mediator
case (\(k = 1\)), where there is only one indirect pathway.

For example, consider a treatment whose effect on a risk outcome operates through
two parallel mediators: blood pressure (BP) and inflammation (IF). Suppose the
direct treatment effect is \(\mathrm{DE}=0.10\), the treatment effect via BP is
\(\mathrm{IE}_{BP}=0.60\) (harmful), and the treatment effect via IF is
\(\mathrm{IE}_{IF}=-0.55\) (protective), yielding \(\mathrm{TE}=0.10+0.60-0.55=0.15\).
Although the net effect is only \(0.15\), the combined magnitude of the three
pathways is \(L=0.10+0.60+0.55=1.25\). The corresponding magnitude shares are
\(\lvert\mathrm{DE}\rvert/L=0.08\), \(\mathrm{MMS}_{BP}=0.48\), \(\mathrm{MMS}_{IF}=0.44\),
which sum to one. The degree of suppression is \(\sigma=0.88\).

The conventional pathway-specific proportions provide little useful interpretation
in this setting. For the BP pathway, \(\mathrm{PM}_{BP}=0.60/0.15=4\), whereas for
the IF pathway, \(\mathrm{PM}_{IF}=-0.55/0.15\approx-3.67\). Both fall outside the
usual proportional range. In contrast, \(\mathrm{MMS}_{BP}=0.48\) and
\(\mathrm{MMS}_{IF}=0.44\) indicate that the two indirect pathways account for
nearly equal shares of the combined magnitude, while \(\sigma=0.88\) quantifies
the substantial attenuation of the TE caused by their opposing directions. The
multiple-pathway formulation therefore preserves the interpretation of MMS and
\(\sigma\) while allowing suppression to arise among several direct and indirect
pathways. We next consider these measures for binary outcomes on the ratio scale.

\section{Binary outcomes on the ratio scale}\label{sec-04}

The additive decomposition \(\mathrm{TE}=\mathrm{DE}+\mathrm{IE}\) applies when
effects are measured on a difference scale. For binary outcomes, effects are
commonly expressed on a ratio scale, such as the odds ratio or risk ratio scale,
where the corresponding decomposition is multiplicative:
\begin{equation*}
\mathrm{TE}_{R}=\mathrm{DE}_{R}\times\mathrm{IE}_{R} \;,
\end{equation*}
where \(R\) denotes the ratio scale. Direct application of MMS and the degree of
suppression to these ratios is not appropriate because the null value is \(1\) rather
than \(0\), and the pathway effects combine multiplicatively rather than additively.
Taking logarithms resolves both issues. The null value \(1\) is mapped to \(0\), and
the multiplicative decomposition becomes additive:
\begin{equation*}
\log(\mathrm{TE}_{R}) = \log(\mathrm{DE}_{R}) + \log(\mathrm{IE}_{R}) \;.
\end{equation*}
Thus, MMS and the degree of suppression can be applied to the corresponding
log-ratio effects using the definitions developed in Sections \ref{sec-02} and
\ref{sec-03}. For a mediation system with one direct and \(k\ge1\) indirect
pathways, let \(L_R\) denote the combined magnitude on the log-ratio scale,
assumed to be positive:
\begin{equation*}
L_R=\lvert\log(\mathrm{DE}_{R})\rvert
+\sum_{j=1}^{k}\lvert\log(\mathrm{IE}_{j,R})\rvert
\end{equation*}
Then we can define,
\begin{align*}
&\mathrm{MMS}_i = \frac{\lvert\log(\mathrm{IE}_{i,R})\rvert}{L_R},
\quad i=1,\ldots,k, \\
&\sigma = 1-\frac{\lvert\log(\mathrm{TE}_{R})\rvert}{L_R}.
\end{align*}
Both measures therefore retain their respective bounds and interpretations on
the log-ratio scale. In particular, ratios above and below \(1\) correspond to
positive and negative log-effects, respectively, so opposing directions on the
ratio scale are represented by opposite signs after transformation.

\section{Simulation study}\label{sec-05}

We conducted a simulation study to examine the finite-sample behavior of PM, MMS,
and \(\sigma\) under consistent mediation, inconsistent mediation, and near-perfect
suppression. Our aim is to illustrate the sampling behavior of these measures
under these conditions, rather than to establish the statistical superiority of
MMS or \(\sigma\) over PM. The last scenario was included to show the behavior of
PM when the TE is very small relative to the magnitudes of the DE and IE. We
considered a single mediator with a continuous outcome and with a binary outcome.
A further simulation with three parallel mediators and a continuous outcome is
reported in the Supplementary Material.

\subsection{Simulation design}\label{simulation-design}

In both settings, the treatment was generated as \(X \sim N(0,1)\), and the
mediator was generated according to
\begin{equation*}
M = \gamma X + \varepsilon_M,
\end{equation*}
where \(\varepsilon_M \sim N(0,1)\) was independent of \(X\). We set \(\gamma=0.5\).
The outcome model, given below for each setting, contained an effect \(\delta=0.4\)
of \(X\) and an effect \(\beta\) of \(M\). Therefore, \(\mathrm{DE}=\delta\). We defined
the IE using the product-of-coefficients approach, in which it is calculated as
the product of the treatment-to-mediator and mediator-to-outcome coefficients
\citep{mackinnon2007, mackinnon2008}. This approach was used for both outcome types.
For the binary outcome, it also avoids the limitations of the
difference-in-coefficients method, which can conflate the estimated IE with
changes arising from the non-collapsibility of the odds ratio \citep{rijnhart2025}.
The TE, \(L\), and the true values of PM, MMS, and \(\sigma\) then follow from
their definitions given earlier.

Only \(\beta\) differed across scenarios. We set \(\beta=0.4\) for consistent
mediation, \(\beta=-0.4\) for inconsistent mediation, and \(\beta=-0.76\) for
near-perfect suppression, giving \(\mathrm{IE}=0.20\), \(-0.20\), and \(-0.38\) and
\(\mathrm{TE}=0.60\), \(0.20\), and \(0.02\), respectively. In the last scenario the
TE is small relative to \(L=0.78\). The true values of \((\mathrm{PM},\mathrm{MMS},\sigma)\)
were \((0.33,\,0.33,\,0)\), \((-1,\,0.33,\,0.67)\), and \((-19,\,0.49,\,0.97)\),
respectively.

\textbf{Continuous outcome.} The outcome was generated as
\begin{equation*}
Y = \delta X + \beta M + \varepsilon_Y,
\end{equation*}
where \(\varepsilon_Y\sim N(0,1)\) was independent of \(X\) and \(\varepsilon_M\). Each
dataset contained \(n=500\) observations.

\textbf{Binary outcome.} The outcome followed a Bernoulli distribution,
\begin{equation*}
Y\sim\operatorname{Bernoulli}(p),
\qquad
\operatorname{logit}(p)=i + \delta X + \beta M,
\end{equation*}
The intercept \(i\) was calibrated numerically in each scenario so that the
population marginal prevalence was \(P(Y=1)=10\%\). The DE and IE were defined on
the log-odds scale, so that \(\mathrm{TE}=\mathrm{DE}+\mathrm{IE}\) holds
additively on that scale (Section \ref{sec-04}) and the true values of PM, MMS,
and \(\sigma\) are those given above. Because the outcome was binary and the
target prevalence was \(10\%\), each simulated dataset contained a larger sample
size \((n=5{,}000)\) than in the continuous-outcome setting \((n=500)\) to obtain
comparable estimation precision.

\subsection{Estimation and performance summaries}\label{estimation-and-performance-summaries}

For each scenario, we generated 10,000 datasets. In the continuous-outcome
setting, we fitted the mediator model \(M \sim X\) and the outcome model
\(Y \sim X + M\) using ordinary least squares (OLS). In the binary-outcome
setting, we fitted the mediator model using OLS and the outcome model using
logistic regression. For each fitted dataset, we estimated the DE by
\(\widehat{\mathrm{DE}}=\hat\delta\) and the IE by
\(\widehat{\mathrm{IE}}=\hat\gamma\hat\beta\). We then computed
\begin{equation*}
\widehat{\mathrm{PM}}
=
\frac{\widehat{\mathrm{IE}}}{\widehat{\mathrm{TE}}},
\quad
\widehat{\mathrm{MMS}}
=
\frac{|\widehat{\mathrm{IE}}|}{\hat L},
\quad
\hat\sigma
=
1-\frac{|\widehat{\mathrm{TE}}|}{\hat L}, 
\end{equation*}
where \(\widehat{\mathrm{TE}}=\widehat{\mathrm{DE}}+\widehat{\mathrm{IE}}\), and
\(\hat L=|\widehat{\mathrm{DE}}|+|\widehat{\mathrm{IE}}|\). Because our interest
is in the range of each estimator, we summarize its sampling distribution by
the minimum, first quartile (Q1), median (Q2), third quartile (Q3), and maximum
across the 10,000 replications (Tables \ref{tab:tbl-two-pathway}--\ref{tab:tbl-binary-out}).

All simulations were performed in R version 4.6.1 \citep{R}, with models fitted by
the \texttt{lm} and \texttt{glm} functions of the \texttt{stats} package and the random number
generator initialized with seed 2026 for each simulation setting. The code to
reproduce the simulation results and tables is available at
\url{https://github.com/shafayetShafee/mms_sigma}.

\subsection{Results}\label{results}

\textbf{Single mediator, continuous outcome.} Under consistent mediation,
\(\widehat{\mathrm{PM}}\) and \(\widehat{\mathrm{MMS}}\) were identical
(median \(0.33\), range \(0.17\)--\(0.56\)) and \(\hat\sigma\) was zero throughout
(Table \ref{tab:tbl-two-pathway}). Under inconsistent mediation,
\(\widehat{\mathrm{PM}}\) was centered near its true value of \(-1\)
(median \(-1.00\)) but had a long left tail (minimum \(-10.22\)), whereas
\(\widehat{\mathrm{MMS}}\) (median \(0.33\), range \(0.22\)--\(0.48\)) and \(\hat\sigma\)
(median \(0.67\), range \(0.44\)--\(0.95\)) stayed close to their true values of
\(0.33\) and \(0.67\). The instability of PM was most pronounced under near-perfect
suppression, where the TE (\(0.02\)) is small relative to \(L\) (\(0.78\)):
\(\widehat{\mathrm{PM}}\) ranged from \(-13{,}785.50\) to \(24{,}332.72\)
(median \(-4.38\); true value \(-19\)), and the positive estimates arose in
replications where \(\widehat{\mathrm{TE}}\) was negative. In the same scenario,
\(\widehat{\mathrm{MMS}}\) (median \(0.49\), range \(0.32\)--\(0.64\)) and \(\hat\sigma\)
(median \(0.95\), range \(0.65\)--\(1.00\)) remained within the unit interval with
medians near their true values of \(0.49\) and \(0.97\).

\textbf{Single mediator, binary outcome.} On the log-odds scale the pattern was the
same (Table \ref{tab:tbl-binary-out}). Under inconsistent mediation,
\(\widehat{\mathrm{PM}}\) again had a median of \(-1.00\) but ranged from
\(-8.67\) to \(61.37\), the positive values reflecting replications with a
wrong-signed \(\widehat{\mathrm{TE}}\), while \(\widehat{\mathrm{MMS}}\)
(\(0.23\)--\(0.50\)) and \(\hat\sigma\) (\(0.46\)--\(0.99\)) remained stable. Under
near-perfect suppression, \(\widehat{\mathrm{PM}}\) ranged from \(-107{,}147.82\) to
\(12{,}363.32\) (median \(-5.28\)), whereas \(\widehat{\mathrm{MMS}}\) (median \(0.49\))
and \(\hat\sigma\) (median \(0.95\)) stayed close to \(0.49\) and \(0.97\).

\begin{table}[!htbp]
\centering
\caption{
  \label{tab:tbl-two-pathway}
  Simulated sampling distributions of the estimated proportion mediated 
  $(\widehat{\mathrm{PM}})$, mediated magnitude share 
  $(\widehat{\mathrm{MMS}})$, and degree of suppression 
  $(\hat{\sigma})$ for a single mediator and a continuous outcome 
  under consistent mediation, inconsistent mediation, and near-perfect 
  suppression. The true direct effect $(\mathrm{DE} = 0.40)$ is the 
  same in all scenarios. The scenarios differ in the true indirect effect 
  $(\mathrm{IE})$ and, consequently, in the true total effect 
  $(\mathrm{TE})$; these are given in the panel headers. Each row 
  summarizes 10,000 simulated datasets of size 500 by the minimum, first 
  quartile (Q1), median, third quartile (Q3), and maximum of the estimates.
}
\centering
\fontsize{9}{11}\selectfont
\setlength{\tabcolsep}{8pt}
\begin{tabular}[t]{lrrrrr}
\toprule
 & Min & Q1 & Median & Q3 & Max\\
\midrule
\addlinespace[0.6em]
\multicolumn{6}{l}{\textbf{\textit{Consistent mediation}}:\; $\mathrm{IE} = 0.20,\; \mathrm{TE} = 0.60$}\\
\addlinespace[3pt]\hspace{1em}$\widehat{\mathrm{PM}}$ & 0.17 & 0.30 & 0.33 & 0.37 & 0.56\\
\hspace{1em}$\widehat{\mathrm{MMS}}$ & 0.17 & 0.30 & 0.33 & 0.37 & 0.56\\
\hspace{1em}$\hat{\sigma}$ & 0.00 & 0.00 & 0.00 & 0.00 & 0.00\\
\midrule
\addlinespace[0.6em]
\multicolumn{6}{l}{\textbf{\textit{Inconsistent mediation}}:\; $\mathrm{IE} = -0.20,\; \mathrm{TE} = 0.20$}\\
\addlinespace[3pt]\hspace{1em}$\widehat{\mathrm{PM}}$ & $-$10.22 & $-$1.25 & $-$1.00 & $-$0.81 & $-$0.39\\
\hspace{1em}$\widehat{\mathrm{MMS}}$ & 0.22 & 0.31 & 0.33 & 0.36 & 0.48\\
\hspace{1em}$\hat{\sigma}$ & 0.44 & 0.62 & 0.67 & 0.71 & 0.95\\
\midrule
\addlinespace[0.6em]
\multicolumn{6}{l}{\textbf{\textit{Near-perfect suppression}}:\; $\mathrm{IE} = -0.38,\; \mathrm{TE} = 0.02$}\\
\addlinespace[3pt]\hspace{1em}$\widehat{\mathrm{PM}}$ & $-$13\,785.50 & $-$10.67 & $-$4.38 & 8.58 & 24\,332.72\\
\hspace{1em}$\widehat{\mathrm{MMS}}$ & 0.32 & 0.46 & 0.49 & 0.51 & 0.64\\
\hspace{1em}$\hat{\sigma}$ & 0.65 & 0.91 & 0.95 & 0.98 & 1.00\\
\bottomrule
\end{tabular}
\end{table}

\begin{table}[!htbp]
\centering
\caption{
  \label{tab:tbl-binary-out}
  Simulated sampling distributions of the estimated proportion mediated 
  $(\widehat{\mathrm{PM}})$, mediated magnitude share 
  $(\widehat{\mathrm{MMS}})$, and degree of suppression 
  $(\hat{\sigma})$ for a single mediator and a binary outcome under 
  consistent mediation, inconsistent mediation, and near-perfect suppression. 
  Effects and measures are computed on the log-odds scale, and the marginal 
  outcome prevalence is 10\%. The true direct effect 
  $(\mathrm{DE} = 0.40)$ is the same in all scenarios. The scenarios 
  differ in the true indirect effect $(\mathrm{IE})$ and, 
  consequently, in the true total effect $(\mathrm{TE})$; these are 
  given in the panel headers. Each row summarizes 10,000 simulated datasets of 
  size 5,000 by the minimum, first quartile (Q1), median, third quartile (Q3), 
  and maximum of the estimates.
}
\centering
\fontsize{10}{12}\selectfont
\setlength{\tabcolsep}{8pt}
\begin{tabular}[t]{lrrrrr}
\toprule
 & Min & Q1 & Median & Q3 & Max\\
\midrule
\addlinespace[0.6em]
\multicolumn{6}{l}{\textbf{\textit{Consistent mediation}}:\; $\mathrm{IE} = 0.20,\; \mathrm{TE} = 0.60$}\\
\addlinespace[3pt]\hspace{1em}$\widehat{\mathrm{PM}}$ & 0.17 & 0.30 & 0.33 & 0.37 & 0.56\\
\hspace{1em}$\widehat{\mathrm{MMS}}$ & 0.17 & 0.30 & 0.33 & 0.37 & 0.56\\
\hspace{1em}$\hat{\sigma}$ & 0.00 & 0.00 & 0.00 & 0.00 & 0.00\\
\midrule
\addlinespace[0.6em]
\multicolumn{6}{l}{\textbf{\textit{Inconsistent mediation}}:\; $\mathrm{IE} = -0.20,\; \mathrm{TE} = 0.20$}\\
\addlinespace[3pt]\hspace{1em}$\widehat{\mathrm{PM}}$ & $-$8.67 & $-$1.22 & $-$1.00 & $-$0.84 & 61.37\\
\hspace{1em}$\widehat{\mathrm{MMS}}$ & 0.23 & 0.31 & 0.33 & 0.35 & 0.50\\
\hspace{1em}$\hat{\sigma}$ & 0.46 & 0.63 & 0.67 & 0.71 & 0.99\\
\midrule
\addlinespace[0.6em]
\multicolumn{6}{l}{\textbf{\textit{Near-perfect suppression}}:\; $\mathrm{IE} = -0.38,\; \mathrm{TE} = 0.02$}\\
\addlinespace[3pt]\hspace{1em}$\widehat{\mathrm{PM}}$ & $-$107\,147.82 & $-$11.83 & $-$5.28 & 8.48 & 12\,363.32\\
\hspace{1em}$\widehat{\mathrm{MMS}}$ & 0.39 & 0.47 & 0.49 & 0.51 & 0.63\\
\hspace{1em}$\hat{\sigma}$ & 0.74 & 0.92 & 0.95 & 0.98 & 1.00\\
\bottomrule
\end{tabular}
\end{table}

\textbf{Summary.} Across the two outcome settings, the measures behaved as their
definitions imply. Under consistent mediation, PM and MMS coincided and
\(\sigma\) was zero. In this setting, PM retains its usual proportional
interpretation and therefore, remains the natural summary. Under inconsistent
mediation, PM took values outside the unit interval and became highly dispersed
as the TE approached zero: under near-perfect suppression its estimates spanned
several orders of magnitude and sometimes had the opposite sign to the true PM.
In all scenarios, \(\widehat{\mathrm{MMS}}\) and \(\hat\sigma\) remained within
\([0,1]\), with medians close to their true values.

These bounds follow from the definitions and do not imply unbiasedness or low
sampling variability; for example, the median of \(\hat\sigma\) was below its true
value under near-perfect suppression. The simulations are illustrative rather
than exhaustive. They show that PM alone can be an unstable summary when the TE
is small relative to the pathway magnitudes, which supports reporting MMS and
\(\sigma\) alongside it when inconsistent mediation is plausible.

\section{Application to the ATLAS program}\label{sec-06}

We now return to the ATLAS example introduced in Section \ref{sec-01}, where the
estimated effects were \(\mathrm{DE}=-0.181\) and \(\mathrm{IE}=0.042\), giving
\(\mathrm{TE}=-0.139\) \citep{goldberg1996, mackinnon2000}. The opposing signs make
this an instance of inconsistent mediation, and the conventional estimate
\(\mathrm{PM}\approx-0.30\) cannot be read as a share of the TE. The combined
magnitude of the two pathways is \(L=0.181+0.042=0.223\), so that
\begin{equation*}
\mathrm{MMS}=\frac{0.042}{0.223}\approx0.188, 
\qquad
\sigma=1-\frac{0.139}{0.223}\approx0.377 .
\end{equation*}
The indirect pathway therefore accounts for approximately \(19\%\) of the combined
magnitude of the DE and IE, and approximately \(38\%\) of that magnitude is
cancelled by their opposing directions. The pathway through the reasons for
using steroids is small relative to the DE but not negligible: it attenuates the
intervention effect from \(-0.181\) to \(-0.139\). These values are computed from
the published point estimates, so they do not reflect estimation uncertainty.

\section{Discussion}\label{sec-07}

MMS is best viewed as a complement to PM rather than a replacement. Under
consistent mediation, MMS coincides with PM and therefore preserves its
conventional interpretation. Under inconsistent mediation, PM loses its
proportional interpretation, whereas MMS remains bounded and the degree of
suppression quantifies the cancellation that PM cannot convey.

Like PM, the interpretations of MMS and the degree of suppression are not
restricted to associational mediation analysis; the same measures can be applied
in causal mediation analysis. As with any summary based on the DE and IE, their
causal interpretation requires that these effects be correctly identified and
estimated. When they are, MMS describes the relative magnitude of the causal
indirect pathway, and the degree of suppression characterizes the extent to
which opposing causal pathways offset one another.

A few practical considerations apply when reporting these measures. First,
because MMS and \(\sigma\) discard the signs of the pathways, they should be
reported together with the DE and IE. Second, \(L\), \(\sigma\), and MMS are defined
relative to a chosen set of pathways. Splitting a single indirect pathway into
two sub-pathways of opposing sign leaves the TE unchanged but increases both \(L\)
and \(\sigma\), so the pathway partition should be justified substantively and
reported explicitly. Third, as with PM, the values of both measures depend on
the effect scale. For binary outcomes, one should therefore state whether the
measures are computed on the risk-difference or the log-ratio scale and, for
odds ratios, whether marginal or covariate-conditional effects are used.

Interval estimation for these measures is not straightforward. We defined MMS
and \(\sigma\) as population quantities and examined only the sampling
distributions of their estimators. Because of the absolute values, \(\sigma\) is
not differentiable wherever the DE, IE, or TE equals zero, and MMS wherever the
DE or IE equals zero; MMS is also undefined when \(L=0\). Standard delta-method
intervals, which rely on differentiability, may therefore perform poorly near
these points, and alternatives such as the bootstrap merit further study.

In sum, MMS and the degree of suppression provide a simple characterization of
both consistent and inconsistent mediation, and can accompany PM in applied
reporting rather than replace it.

\section*{Authors' contributions}

Conceptualization (MNS), Methodology (MNS), Formal analysis (MNS, SKS),
Writing - Original Draft (MNS, SKS, MHP, BS), Writing - Review and Editing
(SKS, MHP, BS), Visualization (MNS, SKS).

\section*{Conflict of interest}

The authors declare that they have no conflicts of interest.

\section*{Funding}

The authors received no specific funding for this work.

\section*{Data availability statement}

No original empirical data were collected or analyzed in this study. The
simulated datasets were generated using the procedures described in the
manuscript and the random seed reported there. The illustrative example uses
previously published effect estimates \citep{mackinnon2000}. The code to
reproduce the simulation results and tables is publicly available at
\url{https://github.com/shafayetShafee/mms_sigma}.


\end{document}